\input epsf
\magnification=1200
\def\qed{\unskip\kern 6pt\penalty 500\raise -2pt\hbox
{\vrule\vbox to 10pt{\hrule width 4pt\vfill\hrule}\vrule}}
\centerline{NATURAL NONEQUILIBRIUM STATES IN QUANTUM STATISTICAL MECHANICS.}
\bigskip\bigskip
\centerline{by David Ruelle\footnote{*}{IHES.  91440 Bures sur Yvette,
France. $<$ruelle@ihes.fr$>$}.}
\bigskip\bigskip\bigskip\bigskip\noindent
	{\leftskip=2cm\rightskip=2cm\sl Abstract.  A quantum spin system is discussed, where a heat flow between infinite reservoirs takes place in a finite region.  A time dependent force may also be acting.  Our analysis is based on a simple technical assumption concerning the time evolution of infinite quantum spin systems.  This assumption, physically natural but currently proved for few specific systems only, says that quantum information diffuses in space-time in such a way that the time integral of the commutator of local observables converges: $\int dt\,||[B,\alpha^tA]||<\infty$.  In this setup one can define a natural nonequilibrium state.  In the time independent case, this nonequilibrium state retains some of the analyticity which characterizes KMS equilibrium states.  A linear response formula is also obtained which remains true far from equilibrium.  The formalism presented here does not cover situations where (for time independent forces) the time translation invariance and uniqueness of the natural nonequilibrium state are broken.\par}
\vfill\eject
	{\bf 0 Introduction.}
\medskip
	Traditional nonequilibrium statistical mechanics has been focussed on approach to equilibrium (Boltzmann and followers) and on situations close to equilibrium (Onsager reciprocity, Green-Kubo formula).  More recently, a fruitful rigorous study of nonequilibrium steady states for classical systems far from equilibrium has been initiated, using the concept of Gaussian thermostat [6], [11].  Among the results are the Gallavotti-Cohen fluctuation theorem [8], [9], the Dettmann-Morriss pairing rule [2], [16], and a general linear response formula [13] (see also Dorfman [3], Ruelle [14] for reviews).  In the approach just referred to, finite classical systems are driven away from equilibrium by nonhamiltonian forces, and cooled by a Gaussian thermostat.  The more natural approach which uses Hamiltonian forces and infinite heat baths is more difficult, and results there are still preliminary [4], [5].
\medskip
	Compared with the classical theory, quantum statistical mechanics exhibits significant differences: equilibrium states (KMS states) are more intrinsically tied to the dynamics, and the forces are fundamentally Hamiltonian.  In particular, the use of a Gaussian thermostat does not appear feasible.  We are thus led to studying infinite systems with Hamiltonian forces.  Fortunately, the dynamics of infinite quantum spin systems is relatively amenable to study: a $C^*$-algebra ${\cal A}$ is associated with the system, and the time evolution is described by a one-parameter family $(\alpha^t)$ of automorphisms of ${\cal A}$.
\medskip
	The physical situation which we wish to discuss is that of a finite quantum system $\Sigma$ interacting with infinite reservoirs $R_a$, ($a=1,2,\ldots$) in equilibrium at different temperatures, chemical potentials, \dots.  The system $\Sigma$ is also acted upon by a force which may be time dependent:
\medskip
\centerline{\epsfbox{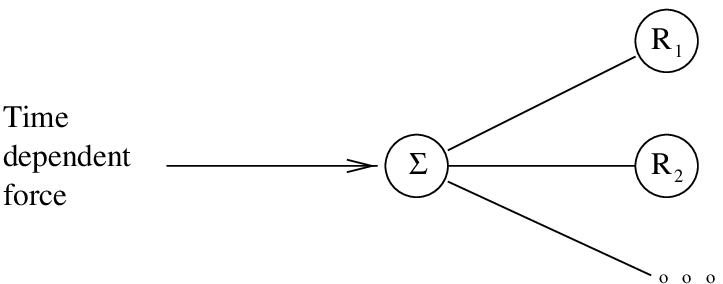}}
\medskip\noindent
For notational purposes it is convenient to write $\Sigma=R_0$.  Note that a finite system subjected to time dependent forces will in general heat up, and that a single reservoir $R_1$ can act as a thermostat.  Another case of interest is when two reservoirs $R_1$, $R_2$ at different temperatures interact via $\Sigma$ (no time dependent force is assumed here).  Variations of the setup just described have been considered by a number of authors (see in particular Hepp and Lieb [10], Spohn and Lebowitz [15], Jak\v si\'c and Pillet [12]); an important feature of the present approach is that it uses more realistic reservoirs.
\medskip
	The strategy of analysis that we shall adopt in this paper will be to compare the time evolution $(\alpha^t)$ of the interacting system described above with a noninteracting time evolution $(\breve\alpha^t)$ where $\Sigma$, $R_1$, $R_2,\ldots$ evolve independently.  Let $\sigma$ be an invariant state for $(\breve\alpha^t)$, where the reservoirs $R_1$, $R_2,\ldots$ are at temperatures $\beta_1^{-1}$, $\beta_2^{-1},\ldots$.  It will be possible to determine a nonequilibrium natural state $\rho_t$ for the interacting system by the condition that it reduces to $\sigma$ in the distant past.
\medskip
	To establish the desired relation between the evolutions $(\alpha^t)$ and $(\breve\alpha^t)$, we assume that the interactions between $\Sigma$ and $R_1$, $R_2,\ldots$, and the time dependent force acting (possibly) on $\Sigma$ are of local nature.  There is however at this point a serious technical problem: our definition of natural nonequilibrium states requires that, when $A$, $B$ are spatially localized, time integrals of the type $\int dt\,||[B,\alpha^tA]||$ converge (and similarly for integrals involving $(\breve\alpha^t)$).  These conditions (further discussed in Section 2.1) are physically natural but have been proved to hold only for very special quantum spin systems.
\medskip
	In view of the above difficulty, we shall in what follows adopt an axiomatic approach.  We shall make technical assumptions {\bf (A1)}-{\bf (A5)} on the dynamics of our quantum spin system, and derive our results from these assumptions.  While {\bf (A1)}-{\bf (A4)} could easily be seen to hold for specific systems (see [1] Section 6.2), there is a problem with {\bf (A5)}, as discussed above.  The interest of the results obtained seems however to justify our axiomatic approach.
\medskip
	In Section 1 we discuss {\bf (A1)}-{\bf (A5)} and derive the existence of a $*$-isomorphism $\omega_t$ between the $C^*$-algebra ${\cal A}$ of the full interacting system, and the $C^*$-algebra ${\cal A}_>$ of the union of the infinite reservoirs $R_1$, $R_2,\ldots$ ($\Sigma$ being omitted).  The isomorphism $\omega_t$ intertwines between the time evolution $(\alpha^t)$ of the full iteracting system and the noninteracting time evolution $(\breve\alpha^t)$ restricted to the union of the reservoirs (see Theorem 1.6).
\medskip
	In Section 2 we show how to define nonequilibrium natural states $\rho_t$ asymptotic in the distant past to noninteracting equilibrium states ($\rho_t$ is related to these states via $\omega_t$).  In Section 3 we consider the case of time independent forces, and assume that the equilibrium states of the noninteracting reservoirs are KMS states at different temperatures.  The nonequilibrium state $\rho_t=\rho$ is now time independent and retains some of the analyticity which characterizes KMS states.
\medskip
	In Section 4 we obtain a linear response formula for $\delta\rho$ when there is a small change $\delta h$ in the interaction of the finite system $\Sigma$ with the infinite reservoirs (the temperature of the reservoirs is not changed).  This quantum linear response formula holds far from equilibrium and is very similar to the corresponding formula for classical systems [13].
\medskip
	{\bf Acknowledgements.}
\medskip
	I am very thankful to both Derek Robinson and Marinus Winnink for enlightening remarks on an earlier version of this paper.
\medskip
	{\bf 1 Assumptions (A).}
\medskip
	The following assumptions {\bf (A1)}-{\bf (A5)} specify our mathematical setup.
\medskip
	{\bf (A1)} {\sl ${\cal A}$, and ${\cal A}_0$, ${\cal A}_1$, ${\cal A}_2$, \dots are (finitely many) $C^*$-algebras with unit elements such that ${\cal A}$ is the $C^*$ tensor product of the ${\cal A}_a$ ($a=0,1,2,\ldots$), and ${\cal A}_0$ is the algebra of $n\times n$ matrices for some finite $n\ge0$.}
\medskip
	There may be several norms on the tensor product $\otimes_{a\ge0}{\cal A}_a$ satisfying the $C^*$ property $||A^*A||=||A||^2$ and the cross-norm property $||\otimes_{a\ge0}A_a||=\prod_{a\ge0}||A_a||$.  If the ${\cal A}_a$ are identified to operator algebras on Hilbert spaces ${\cal H}_a$, the operator norm of $\otimes_{a\ge0}{\cal A}_a$ acting on $\bar\otimes_{a\ge0}{\cal H}_a$ does not depend on the choice of the faithful representations, see [1] Section 2.7.2.  The closure $\bar\otimes_{a\ge0}{\cal A}_a$ of $\otimes_{a\ge0}{\cal A}_a$ for this operator norm is our $C^*$ tensor product of the ${\cal A}_a$.  We shall later also use ${\cal A}_>=\bar\otimes_{a>0}{\cal A}_a$ and denote by ${\bf 1}_0$, ${\bf 1}_>$ the unit element of ${\cal A}_0$, ${\cal A}_>$.
\medskip
	The algebra ${\cal A}_0$ corresponds to the system $\Sigma=R_0$, and ${\cal A}_1$, ${\cal A}_2$, \dots to the reservoirs $R_1$, $R_2$, \dots, while ${\cal A}$ describes the total system.
\medskip
	{\bf (A2)} {\sl $(\breve\alpha^t)$, $(\alpha^t)$ are one-parameter families of $*$-automorphisms of ${\cal A}$ such that
$$	\breve\alpha^t=\otimes_{a\ge0}\,\breve\alpha_a^t      $$
and $(\breve\alpha_a^t)$ is a one-parameter {\rm group} of $*$-automorphisms of ${\cal A}_a$, {\it i.e.}, $\breve\alpha_a^0={\rm identity}$, $\breve\alpha_a^s\breve\alpha_a^t=\breve\alpha_a^{s+t}$, $(\breve\alpha_a^tA_a)^*=\breve\alpha_a^tA_a^*$.}
\medskip
	We shall write $\breve\alpha(t,s)=\breve\alpha^{s-t}$, $\alpha(t,s)=(\alpha^t)^{-1}\alpha^s$, and also $\breve\alpha_>(t)=\breve\alpha_>^t=\otimes_{a>0}\,\breve\alpha_a^t$.
\medskip
	The evolution $(\breve\alpha^t)$ describes the uncoupled systems $\Sigma=R_0$, $R_1$, $R_2$, \dots, while $(\alpha^t)$ describes the total system where $\Sigma$ is coupled to the reservoirs and subjected to time-dependent forces.
\medskip
	{\bf (A3)} {\sl There is a dense subset ${\cal D}\subset{\cal A}$ and for each $s\in{\bf R}$ there is $\epsilon>0$ such that, if $D\in{\cal D}$, the functions 
$$	t'\mapsto\breve\alpha(t',t)\alpha(t,s)D\quad,\qquad
	t'\mapsto\alpha(t',t)\breve\alpha(t,s)D      $$
are (norm-)differentiable when $t,t'\in(s-\epsilon,s+\epsilon)$.}
\medskip
	Clearly, one can assume that ${\cal D}$ is a $*$-subalgebra of ${\cal A}$.  Note also that the families $(\breve\alpha^t)$, $(\alpha^t)$ are strongly continuous because for each $s\in{\bf R}$ and $A\in{\cal A}$, the functions $t\mapsto\breve\alpha^tA$, $\alpha^tA$, or equivalently $t\mapsto\breve\alpha^{-t}A$, $\alpha^{-t}A$, or equivalently for each $B\in{\cal A}$ the functions $t\mapsto\breve\alpha(t,s)B$, $\alpha(t,s)B$ are continuous on $(s-\epsilon,s+\epsilon)$ as uniform limits of continuous (in fact differentiable) functions $t\mapsto\breve\alpha(t-s)D$, $\alpha(t,s)D$ with $D\in{\cal D}$.
\medskip
	{\bf (A4)} {\sl There is a finite dimensional linear space ${\cal F}$ such that 
$$	{\cal A}_0\otimes{\bf 1}_>\subset{\cal F}\subset{\cal A}      $$
and a function $h:{\bf R}\to{\cal F}$ such that $h$ is bounded continuous, self-adjoint ($h=h^*$), and with the notation of {\bf (A3)}
$$   {d\over dt'}(\alpha(t',t)D'-\breve\alpha(t',t)D')|_{t'=t}=-i[h(t),D']   $$
if}
$$	t\in(s-\epsilon,s+\epsilon)\qquad,
	\qquad D'\in\breve\alpha(t,s){\cal D}\cup\alpha(t,s){\cal D}      $$
\medskip
	This expresses that the interaction between the system $\Sigma$ and the reservoirs is of {\it local} nature.
\medskip\goodbreak
	{\bf 1.1 Lemma}
\medskip
	{\sl If $A\in{\cal A}$, the functions
$$	t\mapsto\breve\alpha^t(\alpha^t)^{-1}A\,,\,
	\alpha^t\breve\alpha^{-t}A      $$
are differentiable with derivatives
$$	t\mapsto -i\breve\alpha^t[h(t),(\alpha^t)^{-1}A]\,,\,
	i\alpha^t[h(t),\breve\alpha^{-t}A]      $$}
\medskip
	We first take $A=\alpha^s D$, $D\in {\cal D}$.  Then {\bf (A3)} implies that the function $t\mapsto\breve\alpha^t(\alpha^t)^{-1}A$ is differentiable for $t\in(s-\epsilon,s+\epsilon)$ because
$$	{1\over t'-t}(\breve\alpha^{t'}(\alpha^{t'})^{-1}A
	-\breve\alpha^t(\alpha^t)^{-1}A)=\breve\alpha^{t'}{1\over t'-t}
	(\alpha(t',t)-\breve\alpha(t',t))\alpha(t,s)D      $$
Using {\bf (A4)} we see that the derivative is 
$$	-i\breve\alpha^t[h(t),\alpha(t,s)D]
	=-i\breve\alpha^t[h(t),(\alpha^t)^{-1}A]      $$
so that
$$	\breve\alpha^t(\alpha^t)^{-1}A=\breve\alpha^s(\alpha^s)^{-1}A
	-i\int_s^tdu\,\breve\alpha^u[h(u),(\alpha^u)^{-1}A]\eqno{(1)}      $$
and this formula remains true for all $A\in{\cal A}$ and all $s,t\in{\bf R}$ [cut $[s,t]$ into small subintervals and use the density of ${\cal D}$ in ${\cal A}$].  Since $u\to\breve\alpha^u[h(u),(\alpha^u)^{-1}A]$ is continuous, we see that the derivative of $t\mapsto\breve\alpha^t(\alpha^t)^{-1}A$ is $t\mapsto -i\breve\alpha^t[h(t),(\alpha^t)^{-1}A]$.  The case of $t\mapsto\alpha^t\breve\alpha^{-t}A$ is similar, with
$$	\alpha^t\breve\alpha^{-t}A=\alpha^s\breve\alpha^{-s}A
    +i\int_s^tdu\,\alpha^u[h(u),\breve\alpha^{-u}A]\eqno{(2)\qed}    $$
\medskip
	Note also the formulae
$$	{d\over dt}\breve\alpha(s,t)\alpha(t,s)A
	=-i\breve\alpha(s,t)[h(t),\alpha(t,s)A]\eqno{(3)}      $$
$$	{d\over dt}\alpha(s,t)\breve\alpha(t,s)A
	=i\alpha(s,t)[h(t),\breve\alpha(t,s)A]\eqno{(4)}      $$
which follow directly from Lemma 1.1.
\medskip
	{\bf 1.2 Lemma}
\medskip
	{\sl Let $\alpha^\lambda$ be obtained from $\alpha$ by the
replacement $h\to h+\lambda k$.  Then the functions
$\lambda\to\alpha^\lambda(t,s)A$ are differentiable and}
$$	{d\over d\lambda}\alpha^\lambda(s,t)A
	=i\int_s^td\tau\,\alpha^\lambda(s,\tau)
	[k(\tau),\alpha^\lambda(\tau,t)A]      $$
\indent
	Writing $\Delta\lambda=\lambda'-\lambda$ and
$\Delta\alpha=\alpha^{\lambda'}-\alpha^\lambda$, we obtain from (4)
$$	{d\over dt}\Delta\alpha(s,t)\breve\alpha(t,s)A/\Delta\lambda 
=i\Delta\alpha(s,t)[h(t),\breve\alpha(t,s)A]/\Delta\lambda 
	+i\alpha^{\lambda'}(s,t)[k(t),\breve\alpha(t,s)A]      $$
and using again (4) this is
$$	=\Delta\alpha(s,t)\alpha^\lambda(t,s)/\Delta\lambda
	{d\over dt}\alpha^\lambda(s,t)\breve\alpha(t,s)A
	+i\alpha^{\lambda'}(s,t)[k(t),\breve\alpha(t,s)A]      $$
Writing $\Phi(s,t)=\alpha^\lambda(s,t)\breve\alpha(t,s)$,
$\Phi'_\Delta(s,t)=\Delta\alpha(s,t)\breve\alpha(t,s)/\Delta\lambda$ we
have thus
$$	({d\over dt}\Phi'_\Delta(s,t))\Phi(s,t)^{-1}\Phi(s,t)A
	=\Phi'_\Delta(s,t)\Phi(s,t)^{-1}{d\over dt}\Phi(s,t)A
	+i\alpha^{\lambda'}(s,t)[k(t),\breve\alpha(t,s)A]      $$
hence
$$	({d\over dt}\Phi'_\Delta(s,t))\Phi(s,t)^{-1}\Phi(s,t)A
	+\Phi'_\Delta(s,t)({d\over dt}\Phi(s,t)^{-1})\Phi(s,t)A
	=i\alpha^{\lambda'}(s,t)[k(t),\breve\alpha(t,s)A]      $$
hence
$$	{d\over dt}(\Phi'_\Delta(s,t))\Phi(s,t)^{-1})\Phi(s,t)A
	=i\alpha^{\lambda'}(s,t)[k(t),\breve\alpha(t,s)A]      $$
hence
$$	(\alpha^{\lambda'}(s,t)\alpha^\lambda(t,s)A-A)/\Delta\lambda
	=(\Phi'_\Delta(s,t))\Phi(s,t)^{-1}A      $$
$$	=i\int_s^td\tau\,\alpha^{\lambda'}(s,\tau)
	[k(\tau),\alpha^\lambda(\tau,s)A]\eqno{(5)}      $$
From this it readily follows that $\Phi'_\Delta(s,t)A$ has a limit
$\Phi'(s,t)A$ when $\Delta\lambda\to0$, so that the derivative ${d\over
d\lambda}\alpha^\lambda(s,t)A$ exists (and is equal to
$\Phi'(s,t)\breve\alpha(s,t)A$).  Thus 
$$	({d\over d\lambda}\alpha^\lambda(s,t))\alpha^\lambda(t,s)A
=i\int_s^td\tau\,\alpha^\lambda(s,\tau)[k(\tau),\alpha^\lambda(\tau,s)A]    $$
or
$$	{d\over d\lambda}\alpha^\lambda(s,t)A
=i\int_s^td\tau\,\alpha^\lambda(s,\tau)[k(\tau),\alpha^\lambda(\tau,t)A]
 $$
as announced.\qed
\medskip
	Our last assumption will play a crucial role.
\medskip
	{\bf (A5)} {\sl There are dense subsets  ${\cal E}\subset{\cal A}$ and ${\cal E}_>\subset{\cal A}_>$ such that, if $F\in{\cal F}$, $E\in{\cal E}$, $E_>\in{\cal E}_>$, then
$$	\int_{-\infty}^0ds||[F,(\alpha^s)^{-1}E]||<\infty      $$
$$	\int_{-\infty}^0ds
	||[F,\breve\alpha^{-s}({\bf 1}_0\otimes E_>)]||<\infty      $$}
\medskip
	We may take for  ${\cal E}$ (resp. ${\cal E}_>$) a $*$-subalgebra of ${\cal A}$ (resp. ${\cal A}_>$).  Condition {\bf (A5)} may be taken to mean that, as $s\to-\infty$, $\alpha^sE$, $\breve\alpha^s({\bf 1}_0\otimes E_>)$ diffuse rapidly away from a bounded region, in particular the region of interaction between the system $\Sigma$ and the reservoirs (see the further discussion in section 2.1).
\medskip
	{\bf 1.3 Proposition.}
\medskip
	{\sl There are $*$-morphisms $\omega_t^+:{\cal A}\to{\cal A}$ and $\omega_t^-:{\cal A}_>\to{\cal A}$ such that, for all $A\in{\cal A}$, $A_>\in{\cal A}_>$,
$$	\lim_{s\to-\infty}\breve\alpha(t,s)\alpha(s,t)A=\omega_t^+A      $$
$$	\lim_{s\to-\infty}\alpha(t,s)\breve\alpha(s,t)({\bf 1}_0\otimes A_>)
	=\omega_t^-A_>      $$}
\medskip
	Using (1) and (2) we see that 
$$	\breve\alpha^s(\alpha^s)^{-1}A=(\alpha^0)^{-1}A
	+i\int_s^0dt\,\breve\alpha^t[h(t),(\alpha^t)^{-1}A]      $$
$$	\alpha^s\breve\alpha^{-s}({\bf 1}_0\otimes A_>)
	=\alpha^0({\bf 1}_0\otimes A_>)
  -i\int_s^0dt\,\alpha^t[h(t),\breve\alpha^{-t}({\bf 1}_0\otimes A_>)]  $$
Since $h(\sigma)$ is bounded and takes values in the finite dimensional space ${\cal F}$, {\bf (A5)} shows that the right-hand sides converge when $s\to-\infty$ provided $A=E\in{\cal E}$, $A_>=E_>\in{\cal E}_>$.  By density of ${\cal E}$ in ${\cal A}$ and ${\cal E}_>$ in ${\cal A}_>$, the limits 
$$	\omega_0^+A=\lim_{s\to-\infty}\breve\alpha^s(\alpha^s)^{-1}A\qquad,
\qquad\omega_0^-A_>
=\lim_{s\to-\infty}\alpha^s\breve\alpha^{-s}({\bf 1}_0\otimes A_>)      $$
exist for all $A\in{\cal A}$, $A_>\in{\cal A}_>$, defining $*$-morphisms $\omega_0^+:{\cal A}\to{\cal A}$ and $\omega_0^-:{\cal A}_>\to{\cal A}$.  Therefore the limits asserted in the Proposition also hold, with $\omega_t^+=\breve\alpha^{-t}\omega_0^+\alpha^t$, $\omega_t^-=(\alpha^t)^{-1}\omega_0^-\breve\alpha_>^t$.\qed
\medskip
	{\bf 1.4 Proposition.}
\medskip
	{\sl If $A\in{\cal A}$, $\omega_t^+A={\bf 1}_0\otimes\omega_tA$ where $\omega_t$ is a $*$-morphism ${\cal A}\to{\cal A}_>$.}
\medskip
	Notice first that if $E\in{\cal E}$, {\bf (A5)} gives
$$	\int_{-\infty}^0ds\,\sup_{X\in{\cal A}_0,||X||\le1}
	||[X\otimes{\bf 1}_>,(\alpha^s)^{-1}E]||<\infty      $$
or
$$	\int_{-\infty}^0ds\,\sup_{Y\in{\cal A}_0,||Y||\le1}
	||[Y\otimes{\bf 1}_>,\breve\alpha^s(\alpha^s)^{-1}E]||<\infty      $$
The derivative of $s\mapsto\breve\alpha^s(\alpha^s)^{-1}E$ has bounded norm in view of Lemma 1.1, and the function 
$$	s\mapsto\sup_{Y\in{\cal A}_0,||Y||\le1}
	||[Y\otimes{\bf 1}_>,\breve\alpha^s(\alpha^s)^{-1}E]||      $$
has thus bounded Lipschitz constant.  Therefore
$$	\lim_{s\to-\infty}\sup_{Y\in{\cal A}_0,||Y||\le1}
	||[Y\otimes{\bf 1}_>,\breve\alpha^s(\alpha^s)^{-1}E]||=0      $$
or
$$	(\forall Y\in{\cal A}_0)\qquad
\lim_{s\to-\infty}[Y\otimes{\bf 1}_>,\breve\alpha^s(\alpha^s)^{-1}E]=0      $$
Putting $\breve\alpha^s(\alpha^s)^{-1}E$ in matrix form $(e_{ij}(s))$ with $i,j\in1,\ldots,n$, we see that $e_{ij}(s)\to0$ if $i\ne j$ and $e_{ii}(s)-e_{jj}(s)\to0$ when $s\to-\infty$.  Writing 
$$	E(s)={1\over n}\sum_{i=1}^ne_{ii}(s)      $$
we have $\lim_{s\to-\infty}||\breve\alpha^s(\alpha^s)^{-1}E-{\bf 1}_0\otimes E(s)||=0$.  For general $A$, we can also write $\breve\alpha^s(\alpha^s)^{-1}A$ in matrix form $(a_{ij}(s)$ and define $A(s)={1\over n}\sum_{i=1}^na_{ii}(s)$.  Approximating $A$ by $E\in{\cal E}$ shows that 
$$   \lim_{s\to-\infty}||\breve\alpha^{-s}\alpha^sA-{\bf 1}_0\otimes A(s)||
	=0      $$
or equivalently that $\omega_t^+A={\bf 1}_0\otimes\omega_tA$ where $\omega_t$ is a $*$-morphism ${\cal A}\to{\cal A}_>$.\qed
\medskip
	{\bf 1.5 Proposition.}
\medskip
	{\sl $\omega_t:{\cal A}\to{\cal A}_>$ and $\omega_t^-:{\cal A}_>\to{\cal A}$ are reciprocal $*$-isomorphisms.}
\medskip
	Choose $t\in{\bf R}$ and $A\in{\cal A}$.  Writing $B=\omega_tA$, $A'=\omega_t^-B$, we can in view of Propositions 1.3, 1.4 choose $S<0$ such that if $s<S$,
$$  ||\breve\alpha(t,s)\alpha(s,t)A-{\bf 1}_0\otimes B||<\epsilon\eqno{(6)}  $$
$$	||\alpha(t,s)\breve\alpha(s,t)({\bf 1}_0\otimes B)-A'||
	<\epsilon\eqno{(7)}      $$
(6) implies 
$$  ||A-\alpha(t,s)\breve\alpha(s,t)({\bf 1}_0\otimes B)||<\epsilon\eqno{(8)}  $$
and (7), (8) give
$$	||A-A'||<2\epsilon      $$
hence $\omega_t^-\omega_tA=A'=A$.  If $A_>\in{\cal A}_>$, a similar argument gives $\omega_t\omega_t^-A_>=A_>$.\qed
\medskip
	We may summarize our findings as follows
\medskip\goodbreak
	{\bf 1.6 Theorem.}
\medskip
	{\sl There is a $*$-isomorphism $\omega_t:{\cal A}\to{\cal A}_>$ such that
$$	\lim_{s\to-\infty}\breve\alpha(t,s)\alpha(s,t)A
	={\bf 1}_0\otimes\omega_tA      $$
$$	\lim_{s\to-\infty}\alpha(t,s)\breve\alpha(s,t)({\bf 1}_0\otimes A_>)
	=\omega_t^{-1}A_>      $$
In particular $\omega_t\alpha(t,\tau)=\breve\alpha_>(t,\tau)\omega_\tau$.}
\medskip
	This follows from Propositions 1.3 and 1.5.\qed
\medskip
	{\bf 2 Nonequilibrium states.}
\medskip
	We shall call natural nonequilibrium states those states which, for the evolution $(\alpha^t)$, reduce in the distant past to equilibrium states for the evolution $(\breve\alpha^t)$.  This definition is possible because in the distant past $(\alpha^t)$ and $(\breve\alpha^t)$ are close to each other as a result of our assumption {\bf (A5)}.  We now discuss further this assumption.
\medskip
	{\bf 2.1 Discussion of assumption (A5).}
\medskip
	Physics suggests that, when $A$, $B$ are spatially localized, the estimate
$$	||[B,\alpha^tA]||\approx t^{-d/2}      $$
typically holds for a $d$-dimensional quantum lattice system.  Specifically, some examples treated by Bratteli and Robinson ([1] 5.4.9 and 6.2.14) conform to this diffusive type of behavior (these examples are however rather special).  When $d\ge3$, the $t^{-d/2}$ estimate implies that 
$$	\int dt\,||[B,\alpha^tA]||<\infty      $$
which is basically our assumption {\bf (A5)}.  Consider now the case of time independent forces, {\it i.e.}, let $(\alpha^t)$ be a one-parameter group of automorphisms.  If we have 
$$	\int_{-\infty}^\infty dt\,||[B,\alpha^tA]||<\infty\qquad
	\hbox{for A, B}\in{\cal E}      $$
where ${\cal E}$ is a norm-dense $*$-subalgebra of ${\cal A}$,  Bratteli and Robinson say that $(\alpha^t)$ is $L^1({\cal E})$ {\it asymptotically abelian} ([1] Definition 5.4.8).  Under this condition they prove the existence of our $*$-morphism $\omega^+$ ([1] Proposition 5.4.10) which they call {\it M\o ller morphism} by analogy with quantum scattering theory.  Bratteli and Robinson point out that the difficulty in proving $L^1$ asymptotic abelianness in particular models is not surprising since the existence of the M\o ller morphism is a form of ergodicity\footnote{*}{I am indebted to Derek Robinson for pointing out Section 5.4 of [1] in connection with our assumption {\bf (A5)}.  For an example of nontrivial study of ergodicity in an infinite system, the reader is referred to Fidaleo and Liverani [7].}.  The approach of [1] has technical advantages over the approach adopted here in Section 1, but our discussion has the interest of applying to time dependent forces (and of being self contained).
\medskip
	As we have said, the assumption {\bf (A5)} means that $\alpha^s E$ or $\breve\alpha^s({\bf 1}\otimes E_>)$ rapidly diffuse away from the region of interaction between the system $\Sigma$ and the reservoirs $R_1$, $R_2,\ldots$  Such a diffusion is possible because the reservoirs are infinite, and more precisely of dimension $\ge3$.  This dimensional restriction is physically not surprising if we think of a macroscopic description of the state of our system by a continuous temperature function $T$ tending to finite values $\beta_1^{-1}$, $\beta_2^{-1},\ldots$ at infinity in the different reservoirs.  In the simplest case $T$ should satisfy the heat equation $\triangle T=0$, but if $d=1$, or $2$ this implies that $T$ is constant or unbounded.  We are thus forced to imagine that our reservoirs have dimension 3 or more.
\medskip
	Physically one expects that (for time independent forces) the time translation invariance and uniqueness of the natural nonequilibrium state may be broken.  The formalism presented here does not cover these situations.  Indeed our natural nonequilibrium state $\rho$ will, under natural assumptions\footnote{*}{The state $\rho$ will turn out to be conjugate to the product $\otimes_{a>0}\sigma_a$ of KMS states describing reservoirs, and (using central decomposition) it is natural to assume that the $\sigma_a$ are factor states.  Therefore $\otimes_{a>0}\sigma_a$ and $\rho$ are factor states.  Assume also the asymptotic abelianness condition
$$	\lim_{t\to\infty}||[A,\alpha^tB]||
	=0\qquad\hbox{for }A,B\in{\cal A}      $$
(this is implied by $L^1$ asymptotic abelianness).  Then the mixing property
$$	\lim_{t\to\infty}|\rho(A\alpha^tB)-\rho(A)\rho(B)|=0      $$
holds (see [1] Example 4.3.24).}, satisfy the mixing property $\lim_{t\to\infty}|\rho(A\alpha^tB)-\rho(A)\rho(B)|=0$ and therefore $\rho$ has no nontrivial decomposition into time invariant or periodic states.  In conclusion, we expect that some but not all situations far from equilibrium are covered by our assumption {\bf (A5)}.
\medskip
	{\bf 2.2 Definition of natural nonequilibrium states.}
\medskip
	Let $\sigma_a$ be a state on ${\cal A}_a$, invariant under the one-parameter group $(\breve\alpha_a^t)$ for $a=0,1,2,\ldots$  We shall later impose that the $\sigma_a$ with $a>0$ satisfy the KMS condition (see below).  The GNS construction gives for each $a$ a Hilbert space ${\cal H}_a$, a representation $\pi_a$ of ${\cal A}_a$ by operators on ${\cal H}_a$, a vector $\Omega_a$ such that
$$	\sigma_a(\cdot)=(\Omega_a,\pi_a(\cdot)\Omega_a)      $$
and one-parameter groups $U_a(\cdot)$ of unitary operators such that
$$	U_a(t)\Omega_a=\Omega_a\qquad,\qquad
	U_a(t)\pi_a(A)U_a(t)^{-1}=\pi_a(\breve\alpha_a^tA)      $$
In particular $\sigma=\otimes_{a\ge0}\sigma_a$ is a $\breve\alpha^t$-invariant state on ${\cal A}$.  We say that the time dependent state $\rho_t$ on ${\cal A}$ is a natural nonequilibrium state (NNES) if it is of the form
$$	\rho_t=\lim_{s\to-\infty}\alpha(s,t)^*\sigma      $$
or
$$	\rho_t(A)=\lim_{s\to-\infty}\sigma(\alpha(s,t)A)
	=\lim_{s\to-\infty}\sigma(\breve\alpha(t,s)\alpha(s,t)A)      $$
$$	=\sigma(\omega_t^+A)=\sigma_>(\omega_tA)      $$
where $\sigma_>=\otimes_{a>0}\sigma_a$.  We may thus write
$$	\rho_t=\omega_t^*\sigma_>      $$
which shows that the NNES $\rho_t$ does not depend on the initial state $\sigma_0$ of the system $\Sigma$.  Our definition gives in particular 
$$	\rho_t(\alpha(t,\tau)A)=\rho_\tau(A)      $$
We also have
$$	\rho_t(A)
=\sigma(A)+i\int_{-\infty}^tdu\,\sigma([h(u),\alpha(u,t)A])\eqno{(9)}    $$
[where we have used the formula $-i\int_s^tdu\,\breve\alpha(t,u)[h(u),\alpha(u,t)A]=A-\breve\alpha(t,s)\alpha(s,t)A$, which follows from (3)].
\medskip
	{\bf 2.3 The KMS condition.}
\medskip
	Let $\beta_a>0$.  The $(\breve\alpha_a^t)$-invariant state $\sigma_a$ satisfies the $\beta_a$-KMS condition if, whenever $A$, $B\in {\cal A}_a$, there is a bounded continuous function $F$ on $\{z:0\le{\rm Im}z\le\beta_a\}$, analytic for $z<{\rm Im}z<\beta_a$ and such that for all real $t$
$$	\sigma_a(B.\breve\alpha^tA)=F(t)\qquad,\qquad
	\sigma_a(\breve\alpha^tA.B)=F(t+i\beta_a)      $$
[The physical meaning of this condition is that $\sigma_a$ is an equilibrium state at temperature $\beta_a^{-1}$].
\medskip
	We shall say that $\rho_t$ is a $\beta$-NNES if it is a NNES associated with $\beta_a$-KMS states $\sigma_a$.  It describes thus nonequilibrium in the presence of reservoirs $R_a$ at various temperatures $\beta_a^{-1}$. \medskip
	We assume for simplicity that the $\pi_a$ are faithful representations [this is natural: it is physically reasonable to assume that the ${\cal A}_a$ have quasi-local structure, with simple local algebras, so that the ${\cal A}_a$ are simple algebras.  See [1] Section 2.6.3].
\medskip
	{\bf 3 Time independent forces and nonequilibrium steady states.}
\medskip
	We shall now consider the situation where the forces acting on the system $\Sigma$ are time independent: $h(t)=h$, $\alpha(s,t)=\alpha^{t-s}$, $\omega_t=\omega$, and the NNES $\rho_t=\rho$ is a {\it nonequilibrium steady state} (NESS).  We have thus 
$$	\omega\alpha^t=\breve\alpha_>^t\omega\qquad,\qquad
	\rho=\omega^*\sigma_>      $$
$$	\rho(A)=\sigma(A)+i\int_{-\infty}^0du\,\sigma([h,\alpha^{-u}A])      $$
$$  \rho(B.\alpha^tA)=\sigma_>(\omega B.\breve\alpha_>^t\omega A)\eqno{(10)}$$
when $A$, $B\in{\cal A}$.  If the $\sigma_a$ are $\beta_a$-KMS states for $a>0$, we shall say that $\rho$ is a $\beta$-NESS.
\medskip
	Consider the elements $B\in{\cal A}_>$ such that
$$	||B||_1=\lim_{\epsilon\to0}\inf\{\sum_j\prod_{a>0}||B_{ja}||:
	||B-\sum_j\otimes_{a>0}B_{ja}||<\epsilon\}      $$
These elements form a $*$-algebra ${\cal A}_>^1$ with norm $||.||_1\ge||.||$
\medskip
	Let ${\cal H}_>=\bar\otimes_{a>0}{\cal H}_a$ and ${\cal B}({\cal H_>})$ be the $*$-algebra of bounded operators on ${\cal H}_>$.  Since the $\pi_a$ are faithful, the map $\otimes_{a>0}\pi_a:\otimes_{a>0}{\cal A}_a\to{\cal B}({\cal H}_>)$ extends to a faithful $*$-represen\-ta\-tion $\pi_>$ of ${\cal A}_>$ by bounded operators on ${\cal H}_>$.
\medskip
	Write ${\bf t}=(t_1,t_2,\ldots)$ and let $\hat\alpha^{\bf t}=\otimes_{a>0}\breve\alpha_a^{t_a}=\breve\alpha(t_1)\otimes\breve\alpha(t_2)\otimes\ldots$, {\it i.e.}, $\hat\alpha^{\bf t}$ is the automorphism of ${\cal A}_>$ such that 
$$	\pi_>(\hat\alpha^{\bf t}A)
	=(\prod_{a>0}U_a(t_a))\pi_>(A)(\prod_{a>0}U_>(t_a))^{-1}     $$
($\hat\alpha^{\bf t}$ is unique because $\pi_>$ is faithful).
\medskip
	{\bf 3.1 Proposition.}
\medskip
	{\sl If $A\in{\cal A}_>$, $B\in{\cal A}_>^1$, there is a complex function ${\bf F}$ of the complex variables $z_a$, continuous and bounded by $||A||.||B||_1$ on $\prod_{a>0}\{z_a:0\le{\rm Im}z_a\le\beta_a\}$, analytic in $\prod_{a>0}\{z_a:0<{\rm Im}z_a<\beta_a\}$, and such that
$$	\sigma_>(B.\hat\alpha^{\bf t}A)={\bf F}({\bf t})\qquad,\qquad
	\sigma_>(\hat\alpha^{\bf t}A.B)={\bf F}({\bf t}+i{\bf\beta})      $$
where $\beta=(\beta_1,\beta_2,\ldots)$.}
\medskip
	Take first $A$, $B$ in the algebraic tensor product $\otimes_{a>0}{\cal A}_a$, {\it viz.},
$$	A=\sum_i\otimes A_{ia}\qquad,\qquad
	B=\sum_j\otimes B_{ja}     $$
Then
$$	\sigma_>(B.\hat\alpha^{\bf t}A)
	=\sum_{ij}\prod_{a>0}\sigma_a(B_{ja}\breve\alpha^{t_a}A_{ia})      $$
extends, by the KMS condition for the $\sigma_a$, to a function ${\bf F}$ bounded and continuous on $\prod_{a>0}\{z_a:0\le{\rm Im}z_a\le\beta_a\}$, and analytic in $\prod_{a>0}\{z_a:0<{\rm Im}z_a<\beta_a\}$.  Using the Cauchy formula in several variables we also see that $|{\bf F}(z_1,z_2,\ldots)|$ has the sup-norm
$$	||{\bf F}||=\max_{\eta_1,\eta_2,\ldots}\sup_{\bf t}
|{\bf F}(t_1+i\eta_1\beta_1,t_2+i\eta_2\beta_2,\ldots)|   $$
where $\eta_1$, $\eta_2$,\dots take the values $0$ or $1$.  Using the KMS condition and separating the indices $a'$ with $\beta_{a'}=0$ from the indices $a''$ with $\beta_{a''}=1$ we find that 
$$    {\bf F}(t_1+i\eta_1\beta_1,t_2+i\eta_2\beta_2,\ldots)
=\sum_{ij}\prod_{a'}\prod_{a''}\sigma_{a'}(B_{ja'}.\breve\alpha^{t_a'}A_{ia'})
	\sigma_{a''}(\breve\alpha^{t_a'}A_{ia'}.B_{ja''})      $$
$$	=\sum_{ij}\sigma(\otimes_{a'}B_{ja'}(\otimes_a\breve\alpha^{t_a}A_{ia})
	\otimes_{a''}B_{ja''})
=\sum_j\sigma(\otimes_{a'}B_{ja'}\hat\alpha^{\bf t}A\otimes_{a''}B_{ja''})      $$
hence
$$	||{\bf F}||\le||A||\sum_j\prod_{a\ge0}||B_{ja}||\le||A||.||B||_1      $$
Using the density of $\otimes_{a>0}{\cal A}_a$ in ${\cal A}_>$ and in ${\cal A}_>^1$ concludes the proof of the proposition.\qed
\medskip
	Note that Proposition 3.1 remains true if one changes the assumptions to $B\in{\cal A}_>$, $A\in{\cal A}_>^1$.
\medskip
	{\bf 3.2 Corollary.}
\medskip
	{\sl There is a dense $*$-subalgebra ${\cal A}^{(1)}$ of ${\cal A}$ such that if $A\in{\cal A}$, $B\in{\cal A}^{(1)}$ or  $B\in{\cal A}$, $A\in{\cal A}^{(1)}$, the function $t\mapsto\rho(B.\alpha^tA)$ extends to a continuous function on $\{z:0\le{\rm Im}z\le\min_a\beta_a\}$, analytic in $\{z:0<{\rm Im}z<\min_a\beta_a\}$.}
\medskip
	Taking ${\cal A}^{(1)}=\omega^{-1}{\cal A}_>^1$, this follows from
(10) and Proposition 3.1.\qed
\medskip
	Under suitable physically reasonable conditions one should be able to take ${\cal A}^{(1)}={\cal A}^1=\{A\in{\cal A}:||A||_1<\infty\}$ where 
$$	||A||_1=\lim_{\epsilon\to0}\inf\{\sum_j\prod_{a\ge0}||A_{ja}||:
	||A-\sum_j\otimes_{a\ge0}A_{ja}||<\epsilon\}      $$
\medskip
	{\bf 3.3 The modular group of $\rho$.}
\medskip
	As pointed out by M. Winnink\footnote{*}{Private communication.}, if $({\cal H}_\rho,\pi_\rho,\Omega_\rho)$ is the cyclic representation associated with $\rho$, then $\Omega_\rho$ is cyclic and separating for $\pi_\rho({\cal A})''$, and therefore a modular group $(\tau^t)$ of automorphisms of $\pi_\rho({\cal A})''$ is defined.  In fact we may write ${\cal H}_\rho={\cal H}_>=\bar\otimes_{a>0}{\cal H}_\sigma$, $\pi_\rho=\pi_>\omega=(\otimes_{a>0}\pi_a)\omega$, $\Omega_\rho=\Omega_>=\otimes_{a>0}\Omega_a$ and 
$$	\tau^t\big|_{\pi_\rho({\cal A})}
	=\pi_\rho\otimes_{a>0}\breve\alpha_a^{-\beta_at}\pi_\rho^{-1}      $$
Therefore the modular group $(\tau_t)$ corresponds asymptotically in each reservoir $R_a$ to the noninteracting evolution $(\breve\alpha_a^t)$ accelerated by the factor $-\beta_a$.
\medskip\goodbreak
	{\bf 4 A general linear response formula.}
\medskip
	{\bf 4.1 Proposition.}
\medskip
	{\sl For a perturbation $\delta h(\cdot)$ of the time dependent
interaction $h(\cdot)$, the time dependent nonequilibrium state
$\rho_\cdot$ satisfies the following linear response formula
$$	\delta\rho_t(A)
=i\int_{-\infty}^td\tau\,\rho_t([\alpha(t,\tau)\delta h(\tau),A])   $$
when $A\in(\alpha^t)^{-1}{\cal E}$.  More precisely, if $k:{\bf R}\to{\cal F}$
is bounded continuous self-adjoint and $\rho_t^\lambda$ is the NNES
corresponding to the interaction $h(\cdot)+\lambda k(\cdot)$, then
$\lambda\mapsto\rho_t^\lambda(A)$ is differentiable at $\lambda=0$ when $A\in(\alpha^t)^{-1}{\cal E}$, and}
$$	{d\over d\lambda}\rho_t^\lambda(A)|_{\lambda=0}
	=i\int_{-\infty}^td\tau\rho_t([\alpha(t,\tau)k(\tau),A])      $$

\indent
	Using (9) we have
$$	\rho_t^\lambda(A)=\sigma(A)
	+i\lim_{s\to-\infty}\int_s^tdu\,
	\sigma([h(u)+\lambda k(u),\alpha^\lambda(u,t)A])\eqno{(11)}      $$
where $\alpha^\lambda$ is obtained from $\alpha$ by the replacement
$h\mapsto h+\lambda k$.  From (5) we get also
$$	\alpha^\lambda(u,t)A-\alpha^0(u,t)A
	=i\int_u^td\tau\,\alpha^\lambda(u,\tau)
	[\lambda k(\tau),\alpha^0(\tau,t)A]\eqno{(12)}      $$
where $\alpha^0=\alpha$.  From (11) we obtain
$$	\rho_t^\lambda(A)-\rho_t^0(A)
	=\lim_{s\to-\infty}(\Delta_1(s)+\Delta_2(s))      $$
where 
$$   \Delta_1(s)=i\int_s^tdu\,\sigma([\lambda k(u),\alpha^0(u,t)A])   $$
$$	\Delta_2(s)=i\int_s^tdu\,
	\sigma([h(u)+\lambda k(u),\alpha^\lambda(u,t)A-\alpha^0(u,t)A])   $$
and, using (12), (3), 
$$	\Delta_2(s)=-\int_s^tdu\int_u^td\tau\,\sigma([h(u)+\lambda k(u)
,\alpha^\lambda(u,\tau)[\lambda k(\tau),\alpha^0(\tau,t)A]])      $$
$$=-i\int_s^td\tau\int_s^\tau du\,{d\over du}\sigma(\alpha^\lambda(u,\tau)
	[\lambda k(\tau),\alpha^0(\tau,t)A])      $$
$$	=-i\int_s^td\tau(\sigma([\lambda k(\tau),\alpha^0(\tau,t)A])
-\sigma(\alpha^\lambda(s,\tau)[\lambda k(\tau),\alpha^0(\tau,t)A]))
$$
so that 
$$	(\rho_t^\lambda(A)-\rho_t^0(A))/\lambda
	=i\lim_{s\to-\infty}\int_s^td\tau\sigma(\alpha^\lambda(s,\tau)
	[k(\tau),\alpha^0(\tau,t)A])\eqno{(13)}      $$
If $A\in(\alpha^t)^{-1}{\cal E}$ we can, given $\epsilon>0$, choose $s_0$ such
that 
$$	\int_{-\infty}^sd\tau\,||[k(\tau),\alpha(\tau,t)A]||<\epsilon      $$
if $s<s_0$.  Therefore (13) implies that
$\lambda\mapsto\rho_t^\lambda(A)$ is differentiable at $0$, with
$$	{d\over d\lambda}\rho_t^\lambda(A)|_{\lambda=0}
	=i\lim_{s\to-\infty}\int_{-\infty}^td\tau
	\sigma(\alpha(s,\tau)[k(\tau),\alpha(\tau,t)A])\eqno{(14)}      $$
$$	=i\lim_{s\to-\infty}\int_{-\infty}^td\tau
	\sigma(\alpha(s,t)[\alpha(t,\tau)k(\tau),A])      $$
and the proposition follows readily.\qed
\medskip
	{\bf 4.2 Corollary.}
\medskip
	{\sl Under the conditions of the Proposition 4.1, for time
independent $h(\cdot)$ and $\rho_\cdot=\rho$, we have}
$$	{d\over d\lambda}\rho_t^\lambda(A)|_{\lambda=0}
	=i\int_{-\infty}^td\tau\,\rho([\alpha(t,\tau)k(\tau),A])      $$
$$  =i\int_o^\infty ds\,\rho([\alpha^{-s}k(t-s),A])  $$
\indent
	{\bf 4.3 Remarks.}
\medskip
	If one can choose ${\cal E}$ independent of $\lambda$ such that 
$$   \int_{-\infty}^0d\tau\,||[k(\tau),\alpha^\lambda(\tau,t)A]||   $$
converges when $A\in{\cal E}$, uniformly for
$\lambda\in(\lambda_1,\lambda_2)$, then $\lambda\to\rho_t^\lambda(A)$ is
continuously differentiable on $(\lambda_1,\lambda_2)$, and
$$	{d\over d\lambda}\rho_t^\lambda(A)
	=i\int_{-\infty}^td\tau\,\rho_t^\lambda
	([\alpha^\lambda(t,\tau)k(\tau),A])      $$
Higher order derivatives are also defined, and 
$$	{d^n\over d\lambda^n}\rho_t^\lambda(A)
=i^nn!{\int\cdots\int}_{\tau_1<\cdots<\tau_n<t}d\tau_n\cdots d\tau_1   $$
$$	\rho_{\tau_1}^\lambda
([k(\tau_1),\alpha^\lambda(\tau_1,\tau_2)[k(\tau_2),\cdots
	[k(\tau_n),\alpha^\lambda(\tau_n,t)A]\cdots]])      $$
as is seen by repeated differentiation of 
$$	{d\over d\lambda}\rho_t^\lambda(A)
	=i\lim_{s\to-\infty}\int_{-\infty}^td\tau\,
\sigma(\alpha^\lambda(s,\tau)[k(\tau),\alpha^\lambda(\tau,t)A])      $$
(obtained by restoring the dependence on $\lambda$ in (14)), and Lemma 1.2.  We may also write
$$	{d^n\over d\lambda^n}\rho_t^\lambda(A)
=i^nn!{\int\cdots\int}_{\tau_1<\cdots<\tau_n<t}d\tau_n\cdots d\tau_1   $$
$$	\rho_t^\lambda([\alpha^\lambda(t,\tau_1)k(\tau_1),
	[\alpha^\lambda(t,\tau_2)k(\tau_2),\cdots
	[\alpha^\lambda(t,\tau_n),A]\cdots]])      $$
\medskip
	For time independent $h(\cdot)$ and $k(\cdot)=k$, we have thus 
$$	{1\over n!}{d^n\over d\lambda^n}\rho^\lambda(A)|_{\lambda=0}
=	i^n\int_0^\infty d\sigma_1\cdots\int_0^\infty d\sigma_n
	\rho^\lambda([k,\alpha^\lambda(-\sigma_1)[k,\cdots
	[k,\alpha^\lambda(-\sigma_n)A]\cdots]])      $$
\vfill\eject
	{\bf References.}
\medskip

[1] O. Bratteli and D.W. Robinson.  {\it Operator algebras and quantum statistical mechanics I, II.}  Springer, New York, 1979-1981.  [There is a 2-nd ed. (1997) of vol. II].

[2] C.P. Dettmann and G.P. Morriss.  ``Proof of Lyapunov exponent pairing for systems at constant kinetic energy.''  Phys. Rev. E {\bf 53},R5541-5544(1996).

[3] J.R. Dorfman.  {\it An introduction to chaos in non-equilibrium statistical mechanics.}  Springer, Berlin, to appear.

[4] J.-P. Eckmann, C.-A. Pillet and L. Rey-Bellet.  ``Non-equilibrium statistical mechanics of anharmonic chains coupled to two heat baths at different temperatures.''  Commun. Math. Phys. {\bf 201},657-697(1999).

[5] J.-P. Eckmann, C.-A. Pillet and L. Rey-Bellet.  ``Entropy production in non-linear, thermally driven Hamiltonian systems.''  J. Statist. Phys. {\bf 95},305-331(1999).

[6] D.J. Evans and G.P. Morriss.  {\it Statistical mechanics of nonequilibrium fluids.}  Academic Press, New York, 1990.

[7] F. Fidaleo and C. Liverani.  ``Ergodic properties for a quantum non linear dynamics.''  Preprint.

[8] G. Gallavotti and E.G.D. Cohen.  ``Dynamical ensembles in nonequilibrium statistical mechanics.''  Phys. Rev. Letters {\bf 74},2694-2697(1995).

[9] G. Gallavotti and E.G.D. Cohen.  ``Dynamical ensembles in stationary states.''  J. Statist. Phys. {\bf 80},931-970(1995).

[10] K. Hepp and E.H. Lieb.  ``Phase transitions in reservoir-driven open systems with applications to lasers and superconductors.''  Helvetica Physica Acta {\bf 46},573-603(1973).

[11] W.G. Hoover.  {\it Molecular dynamics}.  Lecture Notes in Physics {\bf 258}.  Springer, Heidelberg, 1986.

[12] V. Jak\v si\'c and C.-A. Pillet.  ``Spectral theory of thermal relaxation.''  J. Math. Phys. {\bf 38},1757-1780(1997).

[13] D. Ruelle.  ``General linear response formula in statistical mechanics, and the fluctuation-dissipation theorem far from equilibrium.''  Phys. Letters {\bf A 245},220-224\ (1998).

[14] D. Ruelle.  ``Smooth dynamics and new theoretical ideas in nonequilibrium statistical mechanics.''  J. Statist. Phys. {\bf 95},393-468(1999).

[15] H. Spohn and J.L. Lebowitz.  ``Stationary non-equilibrium states of infinite harmonic systems.''  Commun. Math. Phys. {\bf 54},97-120(1977).

[16] M.P. Wojtkowski and C. Liverani.  ``Conformally symplectic dynamics and symmetry of the Lyapunov spectrum.''  Commun. Math. Phys. {\bf 194},47-60(1998).

\end